\documentclass{article}
\usepackage{spconf}
\usepackage{amsmath}
\usepackage{amssymb}
\usepackage{amsfonts}
\usepackage{color}
\usepackage[noadjust]{cite}
\usepackage{graphicx}
\usepackage[labelsep=period]{caption}
\usepackage[utf8]{inputenc}
\usepackage[english]{babel}
\usepackage{algorithm}
\usepackage{algpseudocode}
\usepackage{subcaption}
\usepackage{cleveref}
\usepackage[titletoc,title]{appendix}
\ninept

\algnewcommand{\Inputs}[1]{%
  \State \textbf{Inputs: }\parbox[t]{.8\linewidth}{\raggedright #1}
}
\algnewcommand{\Outputs}[1]{%
  \State \textbf{Outputs: }\parbox[t]{.8\linewidth}{\raggedright #1}
}
\algnewcommand{\Initialize}[1]{%
  \State \textbf{Initialize: }\parbox[t]{.8\linewidth}{\raggedright #1}
}
\title{Flexible Multi-Group Single-Carrier Modulation: Optimal Subcarrier Grouping and Rate Maximization}
\name{Yifei Yang$^{\star \dagger}$, Shuowen Zhang$^{\star}$, Joni Polili Lie$^{\dagger}$, and Rui Zhang$^{\star}$}

\address{$^{\star}$Department of Electrical and Computer Engineering, National University of Singapore, Singapore \\
    $^{\dagger}$Wireline \& Perforating, Halliburton Far East, Singapore\\
    Email:\{yifeiyang,shuowen.zhang\}@u.nus.edu, joni.lie@halliburton.com,elezhang@nus.edu.sg}
\begin{document}


\maketitle
\vspace{-1mm}
\begin{abstract}
\vspace{-2mm}
Orthogonal frequency division multiplexing (OFDM) and single-carrier frequency domain equalization (SC-FDE) are two commonly adopted modulation schemes for frequency-selective channels. Compared to SC-FDE, OFDM generally achieves higher data rate, but at the cost of higher transmit signal peak-to-average power ratio (PAPR) that leads to lower power amplifier efficiency. This paper proposes a new modulation scheme, called flexible multi-group single-carrier (FMG-SC), which encapsulates both OFDM and SC-FDE as special cases, thus achieving more flexible rate-PAPR trade-offs between them. Specifically, a set of frequency subcarriers are flexibly divided into orthogonal groups based on their channel gains, and SC-FDE is applied over each of the groups to send different data streams in parallel. We aim to maximize the achievable sum-rate of all groups by optimizing the subcarrier-group mapping. We propose two low-complexity subcarrier grouping methods and show via simulation that they perform very close to the optimal grouping by exhaustive search. Simulation results also show the effectiveness of the proposed FMG-SC modulation scheme with optimized subcarrier grouping in improving the rate-PAPR trade-off over conventional OFDM and SC-FDE.
\end{abstract}


\vspace{-2.2mm}
\section{Introduction}
\vspace{-2.2mm}
\label{sec:intro}
Multicarrier modulation is a promising technique to meet the growing demand for higher data rate and provide enhanced immunity against multipath interference in broadband communications over frequency-selective channels. Particularly, orthogonal frequency division multiplexing (OFDM) has been adopted in many broadband wireless standards \cite{ieee80216,ieee80220}, such as for the Third Generation Partnership Project Long Term Evolution (3GPP-LTE) downlink \cite{3gpp}. OFDM is well known for its flexibility in channel-adaptive rate and/or power allocation for performance optimization \cite{goldsmith,campello,ofdmpa1}. However, OFDM signals generally have high peak-to-average power ratio (PAPR), which requires costly amplifier at the transmitter. Numerous PAPR reduction techniques for OFDM have been reported in the literature \cite{paprmag,paprclip,paprcomp,paprpts,paprslm,paprtr}, but they generally lead to higher implementation complexity. To resolve this issue, single-carrier modulation with frequency-domain equalization (SC-FDE) \cite{scfde2} at the receiver has been proposed and adopted for LTE uplink to reduce the cost of mobile terminals \cite{3gpp,scfdma2}. Compared with non-adaptive OFDM, SC-FDE has significantly lower PAPR and bit error rate (BER) \cite{scfde}. However, the achievable rate of SC-FDE is dominated by the worst frequency subchannel gain, while SC-FDE with channel-adaptive power control provides higher throughput but at the cost of increased PAPR \cite{scfdepapr}. Moreover, generalized frequency division multiplexing (GFDM) was recently proposed \cite{gfdm}, which includes OFDM and SC-FDE as two special cases. However, GFDM in general has non-orthogonal subcarriers, thus requiring complicated receiver design and significantly higher implementation complexity than both OFDM and SC-FDE.

\begin{figure}[!t]
    \centering
    \captionsetup{justification=centering}
    \includegraphics[width=0.9\linewidth, keepaspectratio]{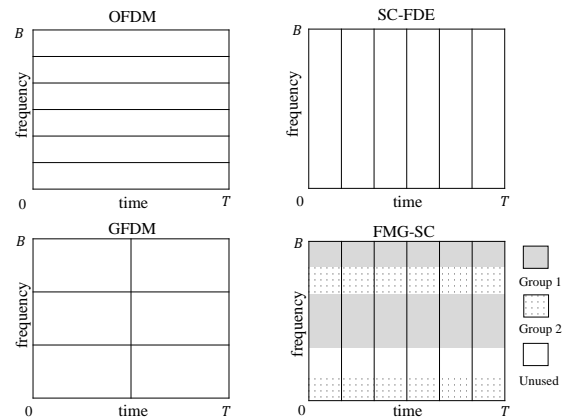}
    \vspace{-3.5mm}
    \caption{Frequency-time representation of different modulation schemes, with $N=6,Q=3,M=2$, and $K=2$.}\vspace{-7mm}
    \label{fig:Fig0}
\end{figure}

Motivated by the limitations of existing modulation techniques, we propose in this paper a new modulation scheme, named flexible multi-group single-carrier (FMG-SC) modulation, which, like GFDM, also encapsulates OFDM and SC-FDE as two special cases, but in a different way. Fig. \ref{fig:Fig0} compares OFDM, SC-FDE, GFDM, and proposed FMG-SC modulated signals using the frequency-time representation. For OFDM, the system total bandwidth $B$ is equally divided by $N$ orthogonal subcarriers, each transmitting a different data symbol at symbol rate $1/T$ in parallel over time, with $T$ denoting the block duration. SC-FDE sequentially transmits $N$ data symbols over $T$ at symbol rate $N/T$, all at the same carrier frequency with bandwidth $B$. GFDM divides the frequency-time dimension into $Q$ subcarriers and $M$ time symbols, where in total $N=QM$ data symbols are transmitted over bandwidth $B$ and in duration $T$. By contrast, FMG-SC transmits $K$ groups of SC-FDE modulated signals simultaneously, where the signals of different groups are orthogonal in frequency as they are modulated by non-overlapping subsets of the $N$ orthogonal subcarriers (see Fig. \ref{fig:Fig0} where in total five data symbols are transmitted in two groups, with three and two symbols in Group 1 and 2, respectively). Notice that for FMG-SC, we consider in general there may be a set of subcarriers that are not used for modulation (e.g., due to deep channel fading), denoted as $\mathcal{S}_0$. Hence, OFDM is a special case of FMG-SC when $K=N$ and $\mathcal{S}_0=\emptyset$, while SC-FDE is another special case with $K=1$ and $\mathcal{S}_0=\emptyset$.

Similar to SC-FDE, the effective receive signal-to-interference-plus-noise ratio (SINR) of each group in FMG-SC is bottlenecked by the subcarrier with the weakest channel gain. Hence, an optimal grouping of subcarriers is crucial to achieving the maximum average sum-rate of all groups in FMG-SC. As the complexity of exhaustively searching over all possible subcarrier groupings is prohibitive, we propose two low-complexity methods to solve this problem approximately. Simulation results show that the proposed methods perform very closely to the optimal exhaustive search in terms of achievable rate for FMG-SC. It is also shown that the proposed FMG-SC modulation with optimized subcarrier grouping can achieve more practically favorable rate-PAPR trade-offs as compared to conventional OFDM and SC-FDE.

\vspace{-2mm}
\section{System Model}
\label{sec:sysmodel}
\vspace{-2mm}
\begin{figure}[!t]
\vspace{-1mm}
    \centering
    \captionsetup{justification=centering}
    \includegraphics[width=\linewidth, keepaspectratio]{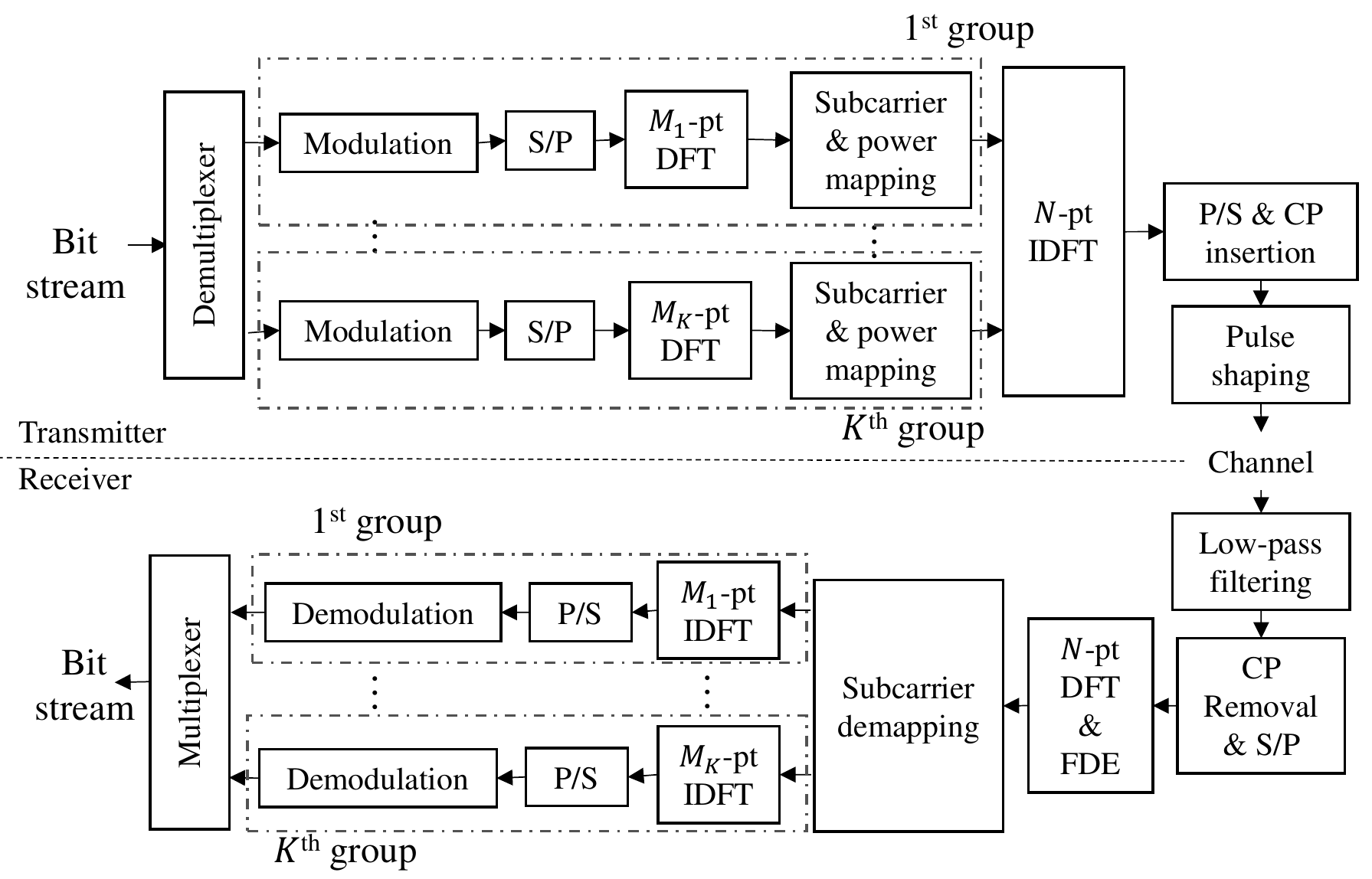}
    \vspace{-5mm}
    \caption{Illustration of the FMG-SC modulation and demodulation.}\vspace{-5mm}
    \label{fig:Fig1}
\end{figure}

We consider point-to-point communication in a frequency-selective channel. Both the transmitter and the receiver are assumed to be equipped with a single antenna. We also assume that perfect channel state information (CSI) is available at the receiver, based on which it determines the parameters needed for designing the transmitter signal modulation and sends them back to the transmitter via a reliable feedback channel. We consider quasi-static channels and for convenience assume that the channel is constant in this paper, if not stated otherwise. Fig. \ref{fig:Fig1} illustrates the proposed FMG-SC modulation scheme, which is described in detail as follows.

Let $\mathcal{N}=\{1,\dotsc,N\}$ denote the set of equally-spaced orthogonal subcarriers, which are further divided into $K$ non-overlapping groups, denoted by the set $\mathcal{K}=\{1,\dotsc,K\}$. Each $k$th group is assumed to consist of $M_k\geq 1$ subcarriers, with $\sum_{k=1}^K M_k\leq N$. Let $\alpha_{k,n}$ indicate whether subcarrier $n$ is allocated to group $k$, $n=1,\dotsc,N$, $k=1,\dotsc,K$, i.e.,
\begin{equation}
\vspace{-1mm}
  \alpha_{k,n}=\begin{cases}
  1, & \mbox{if subcarrier $n$ is assigned to group $k$},  \\
  0, & \mbox{otherwise}.
\end{cases}
\vspace{-1mm}
\label{eqn:subband}
\end{equation}
We then have $\sum_{n=1}^N\alpha_{k,n}=M_k,~\forall k\in\mathcal{K}$. In addition, note that each subcarrier is assigned to at most one group, which yields $\sum_{k=1}^K\alpha_{k,n}\leq 1, \forall n\in\mathcal{N}$. Let $\mathcal{S}_{k}$ denote the set of subcarriers assigned to group $k$ for SC-FDE transmission, which is given by $\mathcal{S}_{k}=\{n|\alpha_{k,n}=1\}$, with $\mathcal{S}_{k}\subseteq\mathcal{N}$. The subcarrier allocation should thus satisfy $\mathcal{S}_{0}\cup\mathcal{S}_{1}\cup\dots\cup\mathcal{S}_{K} = \mathcal{N}$ and $\mathcal{S}_{k}\cap\mathcal{S}_{l}= \emptyset,\forall k\neq l, ~k,l\in\{0\}\cup\mathcal{K}$. Notice that $\mathcal{S}_0$ denotes the set of subcarriers without being assigned to any group $k$, with $|\mathcal{S}_0|\geq 0$.

Let $R$ denote the achievable sum-rate of all groups in bits per second per Hertz (bps/Hz). At the transmitter, all information bits of each transmission block are first demultiplexed into $K$ data streams, each carrying a bit rate $R_k$, where $\sum_{k=1}^KR_k=R$. The $K$ data streams are coded and modulated with SC-FDE modulation \cite{scfde}, and transmitted simultaneously. To reduce the complexity and achieve a low PAPR, we assume equal power allocation among all used subcarriers in all groups (excluding $\mathcal{S}_0$).\footnote{Note that power optimization among subcarriers in different groups can be applied to further improve the achievable rate, but the gain is only marginal as verified by our simulation results when $\mathcal{S}_0$ is properly selected; thus it is not considered in this paper due to the space limitation.} Let $p_{k}$ denote the power allocated to each subcarrier in group $k$. Assume the total transmission power is $P$. The power allocation $p_k$'s should thus satisfy $\sum_{k=1}^K M_kp_{k}\leq P$. With equal power allocation among all used subcarriers, we thus have $p_k=\frac{P}{\sum_{k=1}^{K}\sum_{n=1}^{N}\alpha_{k,n}}$, $\forall k\in\mathcal{K}$.

Without loss of generality, we focus on the transmission of group $k$ only. Let $\boldsymbol{x}\in\mathbb{C}^{M_k\times 1}$ denote the information symbol vector, whose entries are modelled as independent and identically distributed (i.i.d.) random variables each with zero mean and unit variance. By referring to Fig. \ref{fig:Fig1}, the transmitted signal from group $k$ before cyclic prefix (CP) insertion is expressed as
\begin{equation}
\vspace{-1mm}
	\tilde{\boldsymbol{x}}=\boldsymbol{F}_N^H\boldsymbol{A}\boldsymbol{P}\boldsymbol{F}_{M_k}\boldsymbol{x},
\end{equation}
where $\boldsymbol{F}_{M_k}$ is an $M_k\times M_k$ discrete Fourier transform (DFT) matrix, $\boldsymbol{F}_N^H$ is an $N\times N$ inverse DFT (IDFT) matrix, $\boldsymbol{P}\in\mathbb{R}^{M_k\times M_k}_{+}$ is the diagonal power allocation matrix with all diagonal entries given by $\sqrt{p_k}$, and $\boldsymbol{A}\in\mathbb{Z}^{N\times M_k}$ is the subcarrier-group mapping matrix with $A_{i,j}=\{1|\alpha_{k,i}=1,\sum_{n=1}^i \alpha_{k,n}=j\}$ and $A_{i,j}=0$ otherwise, $i=1,\dotsc,N$, $j=1,\dotsc,M_k$. For example, consider the case with $N=4$, $M_k=2$, we have
\begin{equation}\label{eqn:A}
  \boldsymbol{A}= \begin{bmatrix}
1 & 0 & 0 & 0\\
0 & 0 & 1 & 0
\end{bmatrix}^T,
\end{equation}
if the first and third subcarriers are allocated to group $k$. It is worth noting that to enable the above modulation, there are in general two sets of parameters for the receiver to feed back to the transmitter:
\setlength{\itemsep}{1pt}
\begin{enumerate}
	\item Subcarrier-group mapping at each subcarrier: $\{\alpha_{k,n}\}$, $k=1,\dotsc,K$, $n=1,\dotsc,N$;
	\item Rate assignments for each group: $\{R_k\}$, $k=1,\dotsc,K$.
\end{enumerate}
However, for the practical case with a small number of groups, the feedback complexity is moderate and affordable. For example, consider $K=1$, then the feedback parameters are reduced to only the subcarrier indices in $\mathcal{S}_0$ and the transmission rate.

At the receiver, let $\boldsymbol{H}\in\mathbb{C}^{N\times N}$ denote the complex baseband channel in time domain. The use of CP renders $\boldsymbol{H}$ to be a circulant matrix and thus expressed via eigenvalue decomposition as $\boldsymbol{F}_N^H\boldsymbol{\Lambda} \boldsymbol{F}_N$, where $\boldsymbol{\Lambda}\in\mathbb{C}^{N\times N}$ is a diagonal matrix with each $(n,n)$-th entry $h_n$ being the complex channel gain at the $n$th subcarrier in the frequency domain. Let $\boldsymbol{z}\in\mathbb{C}^{N\times 1}$ denote the receiver noise, whose entries are modelled as i.i.d. circularly symmetric complex Gaussian (CSCG) random variables each with zero mean and variance $\sigma^2$. Hence, the received signal after CP removal is expressed as
\begin{equation}\label{eqn:y}
\vspace{-1mm}
  \boldsymbol{y} =\boldsymbol{H}\tilde{\boldsymbol{x}}+\boldsymbol{z}= \boldsymbol{F}_N^H\boldsymbol{\Lambda} \boldsymbol{A}\boldsymbol{P}\boldsymbol{F}_{M_k}\boldsymbol{x}+\boldsymbol{z}.
\end{equation}
As shown in Fig. \ref{fig:Fig1}, an FDE is applied to all subcarriers based on the criterion of minimum mean square error (MMSE) \cite{screc}. The equalized signal of group $k$ in time domain is thus given by
\begin{align}
\vspace{-1mm}
	\tilde{\boldsymbol{y}}=&\boldsymbol{F}_{M_k}^H\boldsymbol{A}^H\boldsymbol{T}\boldsymbol{F}_N\boldsymbol{y}\\
	=&\boldsymbol{F}_{M_k}^H\boldsymbol{A}^H\boldsymbol{T}\boldsymbol{\Lambda} \boldsymbol{A}\boldsymbol{P}\boldsymbol{F}_{M_k}\boldsymbol{x}+\boldsymbol{F}_{M_k}^H\boldsymbol{A}^H\boldsymbol{T}\boldsymbol{F}_N\boldsymbol{z},
\vspace{-1mm}
\end{align}
where $\boldsymbol{T}\in\mathbb{C}^{N\times N}$ is the diagonal MMSE-FDE matrix with the $(n,n)$-th entry denoted by $\frac{\sqrt{p_k}h_n^*}{p_k|h_n|^2+\sigma^2}$, $n=1,\dotsc,N$, where $(\cdot)^*$ represents the complex conjugate operation.
Therefore, the receive SINR of group $k$ can be shown to be \cite{screc}
\begingroup\makeatletter\def\f@size{9}\check@mathfonts
\begin{equation}
	\gamma_{k}(\{\alpha_{k,n}\})
=\frac{1}{\frac{1}{M_k}\sum_{n=1}^{N}\alpha_{k,n}\left(\frac{\sigma^2}{\sigma^2+{p_{k}|h_{n}|^2}}\right)}-1,~k=1,\dotsc,K.\label{eqn:gamma}
\end{equation}
\endgroup
Hence, the achievable rate in bps/Hz at the $k$th group is given by
\begingroup\makeatletter\def\f@size{9}\check@mathfonts
\begin{equation}\label{eqn:rk}
	R_{k}(\{\alpha_{k,n}\})=\frac{M_k}{N}\log_2\left(1+\frac{\gamma_{k}(\{\alpha_{k,n}\})}{\Gamma}\right),
\end{equation}
\endgroup
where $\Gamma \geq 1$ denotes the gap for the achievable rate from the channel capacity owing to the employment of a practical modulation and coding scheme (MCS) \cite{cioffi2}.

\vspace{-2mm}
\section{Problem Formulation}
\vspace{-2mm}
In this paper, we aim to maximize the achievable sum-rate over all groups for our proposed FMG-SC under a given number of groups,  $K\geq 1$, by optimizing the subcarrier-group mapping $\{{\alpha}_{k,n}\}$. Hence, the problem is formulated as

\begin{align}
\vspace{-1mm}
\mathrm{(P1)}:~\mathop{\mathtt{maximize}}_{\{\alpha_{k,n}\}}  &~\sum_{k=1}^{K}R_{k}(\{\alpha_{k,n}\})	\label{eqn:prob0}\\
	\mathtt{subject\; to}
&~~\alpha_{k,n}\in \{0,1\}, \quad \forall k \in \mathcal{K},\; \forall n \in \mathcal{N} \label{eqn:const3}\\
&~~\sum_{k=1}^K\alpha_{k,n}\leq 1, \quad \forall n\in\mathcal{N}.	\label{eqn:const4} \vspace{-1mm}
\end{align}

Note that (P1) is a non-convex combinatorial optimization problem due to the binary constraints on $\{\alpha_{k,n}\}$ in (\ref{eqn:const3}). A direct approach for finding its optimal solution is via searching over all possible subcarrier-group mappings and selecting the one with the maximum sum-rate. Note that for each subcarrier, there are $K+1$ possible group mappings, since it can be assigned to any of the $K$ groups $\{\mathcal{S}_k\}_{k\in\mathcal{K}}$, or to $\mathcal{S}_0$ (i.e., unused). The complexity of computing the sum rate of all groups given any subcarrier-group mapping can be shown to be $\mathcal{O}(NK)$, and that of searching over all possible mappings is $\mathcal{O}\left((K+1)^N\right)$. Hence, the total computational complexity of exhaustive search for the optimal $\{\alpha_{k,n}\}$ is $\mathcal{O}\left(NK(K+1)^{N}\right)$, which is unaffordable for practical systems with large $N$.
\vspace{-2mm}
\section{Proposed Solution}
\label{sec:sol}
\vspace{-2mm}
To avoid the prohibitive complexity of exhaustive search, in this section, we present two suboptimal methods with lower complexity to solve (P1).
\subsection{Set Partitioning Optimal Search (SPOS)}
\label{sec:exsearch}
\vspace{-2mm}
Notice from (\ref{eqn:gamma}) that $(1+\gamma_k)$ is the harmonic mean of $\big\{1\!+\!\frac{p_{k}|h_{n}|^2}{\sigma^2}\big\}_{n\in\mathcal{S}_k}$, where $\frac{p_{k}|h_{n}|^2}{\sigma^2}$ is the signal-to-noise ratio (SNR) of subcarrier $n$ that belongs to group $k$. The harmonic mean operation is known to be dominated by the smallest element in its arguments. As a result, it can be shown that for a group of subcarriers with a minimum subcarrier SNR of $\gamma'_0$, its effective SINR is upper-bounded by $\left(M_k(1+\gamma'_0)-1\right)$. Thus, the achievable rate of each group with SC-FDE transmission is bottlenecked by its worst subcarrier SNR. By intuition, subcarriers with similar SNR values should be grouped together to mitigate the above effect.

Motivated by this, we solve (P1) approximately by considering the problem of finding the optimal partition of $N$ subcarriers sorted in an increasing order of their SNRs into $K+1$ non-overlapping groups to maximize the sum-rate. We thus term this method as set partitioning optimal search (SPOS). Equivalently, we need to determine the locations to insert $K$ bars that separate the $N$ sorted subcarrier SNRs into $K+1$ bands. The group of subcarriers with the lowest SNR values belong to $\mathcal{S}_0$ and thus are not used for transmission, while the transmission power is equally allocated among the rest of the subcarriers. This is equivalent to a \emph{set partitioning problem} in combinatorial optimization. Specifically, there are $N$ locations to place the $K$ bars, denoted by $\boldsymbol{b}=[b_1,\dotsc,b_{K}]^T\in\mathbb{Z}_+^{K}$. Hence, there are $N\choose K$ possible divisions, and exhaustively searching over all of them requires a complexity of $\mathcal{O}\left(N^K\right)$.
Note that each possible $\boldsymbol{b}$ determines a unique set of values for $\{{\alpha}_{k,n}\}$, and the complexity of sorting subcarriers is $\mathcal{O}(N\log N)$. Thus, the overall complexity of the proposed SPOS is $\mathcal{O}\left(N\log N+N^{K+1}K\right)$, which is lower than that of exhaustively searching $\{\alpha_{k,n}\}$ since it usually holds that $K\ll N$.
\vspace{-2mm}
\subsection{Set Partitioning Gradient-based Search (SPGS)}
\vspace{-2mm}
Alternatively, we present a more efficient iterative algorithm that further reduces the complexity of the above SPOS, by leveraging a gradient-based search for solving (P1). Let $\tilde{\boldsymbol{I}}=[\tilde{I}_1, \dotsc, \tilde{I}_N]^T\in\mathcal{N}^N$ denote the sorted subcarrier indices, and $\boldsymbol{g}=[g_1, \dotsc, g_N]^T\in\mathbb{R}_+^N$ denote the corresponding sorted subcarrier SNR values, i.e. $g_m$ is the SNR value of subcarrier $n$ for $n\in\mathcal{N}$ in the original order if $\tilde{I}_m=n$. In each inner iteration, we move $b_k$ at most one position to the direction that maximizes (P1) with other $b_i$'s fixed, $i\neq k,~i,k\in\mathcal{K}$, i.e., $b_k \leftarrow b_k+\delta_k$, where $\delta_k\in\{-1,0,1\}$. This is carried out sequentially over $k=1,\dotsc, K$, which completes an outer iteration. The outer iteration terminates when no change is made to all bar locations. Note that for each $b_k$, its domain is bounded by $[b_{k-1}+1, b_{k+1}-1]$, where $b_0=-1$ and $b_{K+1}=N$.

It is worth noting that since (P1) is not a convex problem, i.e., it may have multiple locally optimal points, the proposed algorithm may terminate at a local optimum. A good choice of the initial $\boldsymbol{b}$ is thus crucial to its performance. One possible way is to randomly generate multiple initial points and choose the one that produces the highest sum-rate after convergence. However, this may be computationally inefficient. Alternatively, as the harmonic mean of a set of positive numbers is dominated by the smallest number in the set, and the subcarriers with similar SNR values should be grouped together, we propose to set the initial set of $\boldsymbol{b}$ as
$\left[\lfloor\frac{N}{K+1}\rfloor,\dotsc,\lfloor\frac{NK}{K+1}\rfloor\right]^T$, which partitions the $N$ sorted subcarriers into $K$ equal-size bands.
The complexity of the above proposed set partitioning gradient-based search (SPGS) depends on the number of outer iterations for convergence, which is difficult to analyze in general. The best case only needs one outer iteration, while the worst case may have the iteration number linear in $N$ as observed from our simulations. Hence, the complexity of SPGS is at most $\mathcal{O}\left(N\log N+N^2K^2\right)$, which is even lower than that of the previous SPOS.

\vspace{-2mm}
\section{Simulation Results}
\label{sec:sim}
\vspace{-2mm}

This section presents simulation results to evaluate the performance of our proposed FMG-SC modulation with different subcarrier grouping methods. The frequency-selective channel is assumed to consist of $L=8$ paths with an exponential power delay profile, where each tap coefficient is modeled as an independent CSCG random variable with zero mean and variance determined by the power delay profile. The total average power of all paths is normalized to unit, and thus the SNR is defined as $P/\left(N\sigma^2\right)$. We set $N=64$, unless stated otherwise. Each simulation result is averaged over 1000 independent channel realizations. The CP length for all modulation schemes considered (OFDM, SC-FDE, and FMG-SC) is set equal to $L$ and the rate loss due to CP is ignored since it is a constant percentage for all schemes.

\begin{figure}[!t]
    \centering
    \captionsetup{justification=centering}
    \includegraphics[width=0.8\linewidth, keepaspectratio]{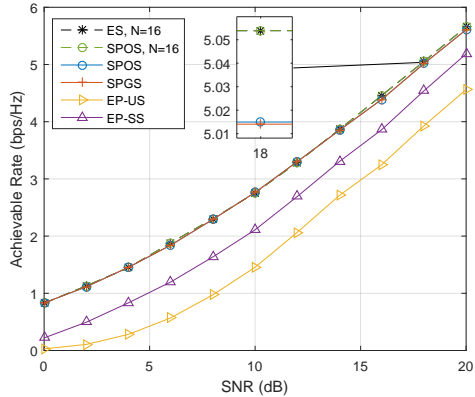}
    \vspace{-2mm}
    \caption{Achievable rate comparison of different subcarrier grouping methods for FMG-SC with $K=2$ and $\Gamma = 1$.}
    \vspace{-3mm}
    \label{fig:algocomp}
\end{figure}

\begin{figure}[!b]
\vspace{-2mm}
    \centering
    \captionsetup{justification=centering}
    \includegraphics[width=0.8\linewidth, keepaspectratio]{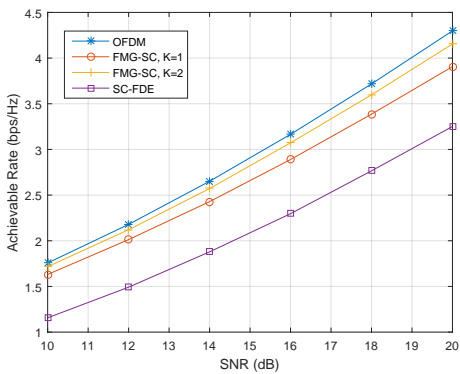}
    \vspace{-2mm}
    \caption{Achievable rate comparison of OFDM, FMG-SC, and SC-FDE under $\Gamma=4.54$dB with 1/3 convolutional code.}
    \label{fig:multi_rate}
\end{figure}

First, we evaluate the performance of our proposed low-complexity subcarrier grouping for FMG-SC. Fig. \ref{fig:algocomp} shows the achievable rates of different grouping methods with $K=2$ and $\Gamma=1$ (0dB). We consider three benchmark schemes, including exhaustive search (ES), equal partition with unsorted subcarriers (EP-US), and equal partition with sorted subcarriers (EP-SS). EP-US divides the subcarrier set $[1,\dotsc,N]$ into $K$ equal bands directly whereas EP-SS operates similarly on sorted subcarriers in the set $\tilde{\boldsymbol{I}}$. Notice that we have assumed all the subcarriers are used in this example, i.e., $\mathcal{S}_0 =\emptyset $. Due to its high computational complexity, we only consider $N=16$ for ES and compare it with the proposed SPOS with $N=16$. From Fig. \ref{fig:algocomp}, it is observed that the proposed SPOS performs as good as the optimal ES. For other results shown in Fig. \ref{fig:algocomp}, we consider $N=64$. It is observed that the proposed SPGS has very close achievable rate compared to the proposed SPOS, although it requires even lower complexity.  Additionally, it is also observed that benchmark scheme ES-US has the lowest achievable rate, while the rate is improved if the subcarriers are sorted based on SNRs as in the benchmark scheme ES-SS. The two proposed subcarrier grouping methods significantly outperform the two benchmark schemes with equal partition, especially at low SNR values.

\begin{figure}[!t]
    \centering
    \captionsetup{justification=centering}
    \includegraphics[width=0.8\linewidth, keepaspectratio]{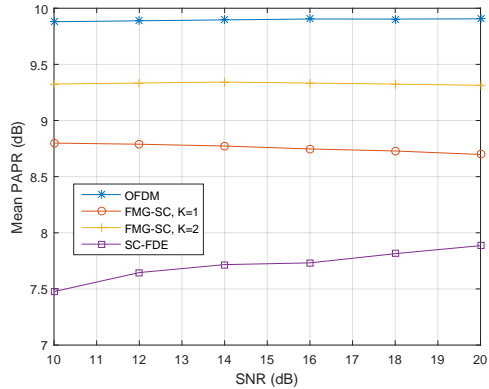}
    \vspace{-2mm}
    \caption{Mean PAPR comparison of OFDM, FMG-SC, and SC-FDE under $\Gamma=4.54$dB with 1/3 convolutional code.}
    \vspace{-3mm}
    \label{fig:multi_papr}
\end{figure}

Fig. \ref{fig:multi_rate} and Fig. \ref{fig:multi_papr} compare the achievable rate and the mean PAPR of different modulation schemes, respectively. Specifically, OFDM with the optimal water-filling (WF) based power and bit allocation over all subcarriers is considered, while for SC-FDE and the proposed FMG-SC with $K=1$ or 2 groups, equal power allocation over all/used subcarriers is assumed. For FMG-SC, the proposed SPGS subcarrier grouping method is assumed, since it yields nearly optimal performance with the lowest complexity, as shown in Fig. \ref{fig:algocomp}. We assume that a rate-1/3 convolutional code with constraint length of 3 is used, for which the SNR gap is $\Gamma=4.54$dB at a target BER of $10^{-6}$  \cite{proakis}. With practical modulation and coding, the achievable rate needs to be discrete for each channel realization. Thus, we consider $M$-QAM modulation with bit granularity of 1/3. Root-raised cosine (RRC) pulse shaping function with rolloff factor of 0.1 is also assumed. From Fig. \ref{fig:multi_rate}, it is observed that there are significant rate gains by using FMG-SC even with $K=1$ (i.e., single group) as compared to SC-FDE. This indicates that a simple transmit adaptation by nulling the weakest subcarriers in SC-FDE leads to significant improvement in achievable rate. With increased $K$ (e.g., $K=2$), the achievable rate of FMG-SC approaches that of WF-OFDM. However, as observed in Fig. \ref{fig:multi_papr}, the improvement in data rate by FMG-SC over SC-FDE is at the cost of increased mean PAPR, which is more pronounced when $K$ increases from 1 to 2 in FMG-SC. Hence, the proposed FMG-SC modulation with optimized subcarrier grouping provides more flexible rate-PAPR trade-offs between conventional SC-FDE and OFDM, which are practically appealing.

\vspace{-2mm}
\section{Conclusion}
\vspace{-2mm}
This paper proposes a new general modulation scheme termed FMG-SC for broadband communication over frequency-selective channels, which encapsulates conventional OFDM and SC-FDE modulations as special cases. We study the optimal subcarrier grouping for FMG-SC to maximize the achievable rate, and propose two low-complexity methods that can find nearly optimal solutions efficiently. It is shown by simulation that the proposed FMG-SC modulation with optimized subcarrier grouping has achievable rate close to that of WF-OFDM yet with lower PAPR. Meanwhile, its achievable rate significantly outperforms that of SC-FDE at the cost of moderately higher PAPR. Hence, the proposed FMG-SC provides more practically favorable rate-PAPR trade-offs over existing OFDM and SC-FDE.

\bibliographystyle{IEEEtran}
\bibliography{fmgsc_conf}

\end{document}